# Direct Observation of Intravalley Phonon Scattering of 2s Excitons in MoSe$_2$ and WSe$_2$ Monolayers

*Liam P. McDonnell [1], Jacob Viner [1], Pasqual Rivera [2], Xiaodong. Xu [2], David C. Smith [1*]*

1 School of Physics and Astronomy, University of Southampton, Southampton SO17 1BJ, United Kingdom.

2 Department of Physics, University of Washington, Seattle, WA, USA

We present a high-resolution resonance Raman study of hBN encapsulated MoSe$_2$ and WSe$_2$ monolayers at 4 K using excitation energies from 1.6 eV to 2.25 eV. We report resonances with the WSe$_2$ A2s and MoSe$_2$ A2s and B2s excited Rydberg states despite their low oscillator strength. When resonant with the 2s states we identify new Raman peaks which are associated with intravalley scattering between different Rydberg states via optical phonons. By calibrating the Raman scattering efficiency and separately constraining the electric dipole matrix elements, we reveal that the scattering rates for k=0 optical phonons are comparable for both 1s and 2s states despite differences in the envelope functions. We also observe multiple new dispersive Raman peaks including a peak at the WSe$_2$ A2s resonance that demonstrates non-linear dispersion and peak-splitting behavior that suggests that the dispersion relations for dark excitonic states at energies near the 2s state are extremely complex.

**Main Body**

Many of the proposed applications of the monolayer TMDCs and their heterostructures are based upon their strongly bound, high oscillator strength excitonic and trionic states [1,2]. Understanding the physics of these states is central to delivering these applications. Valleytronics relies on the suppression of intervalley scattering[3]. Polaritonics [4,5] requires understanding the separation of the dipole and phonon components of the transition linewidths [6]. Quantum computing could take advantage of the ability to generate coherent populations using Raman processes [7]. In addition, understanding the behavior of heterostructures, in which intralayer excitons at different energies interact with each other and interlayer excitons [8], will require a thorough understanding of excitonic physics in the constituent monolayers.

There has been significant progress in our understanding of excitons in TMDCs in the last few years [3,9–12]. The existence and energies of a wide range of excitonic states are now well established [7,11,13,14]. After starting with the three main excitonic bands, A, B and C [9,15], attention is now focused on the excited excitonic states including both the bright s states [12,16] and the two photon accessible p states [17,18]. The effect of the envelope function on the oscillator strength of the excitons has been found to be in line with expectation [19,20]. There are predictions that the B excitons should have enhanced oscillator strength relative to the A excitons because of exchange interactions [21] although only limited experimental validation [22]. Attention is starting to turn to understanding the scattering of the excited state excitons with measurements of the temperature dependence of the transition linewidths and attempts to model the total phonon scattering using simplified excitonic band structures [23,24]. The importance of a wide variety of dark excitonic states is now well established [25] however the difficulties of experimentally accessing these states and calculating full excitonic dispersion relations is hampering fully understanding them.

Raman scattering, and particularly resonance Raman spectroscopy, has a proven track record in elucidating the physics of excitons and phonons and their interaction in the TMDCs [26–29]. Its ability to resolve the phonons involved in excitonic scattering has led to discoveries such as the coupling of substrate phonons to excitons in $WSe_2$ monolayers [30]. The ability to directly probe dark states via multi-phonon processes has elucidated the importance of intervalley scattering and allowed mechanisms for exciton-trion scattering via dark states to be explored [31,32]. However, Raman scattering still has a lot more to give to our understanding of the fundamental physics of excitons in TMDCs. In particular, the only reports of resonance Raman with excited states in monolayer TMDCS are via hBN encapsulation phonons[30] and there are no reports of resonance with B2s excitons.

In this paper we discuss the results of resonance Raman experiments performed on high quality hBN encapsulated monolayers of $WSe_2$ and $MoSe_2$ at 4 K along with photoluminescence and reflectivity spectra taken at the same location on the samples. Whilst the data presented in the main body of this paper is from the best quality samples all key observations are supported by measurements on at least one other monolayer (see SI). As shown in Fig 1, we observe Raman resonances associated with the A1s, A2s, B1s and B2s excitonic states in $MoSe_2$ and the A1s, A2s, B1s excitonic states in $WSe_2$; the $WSe_2$ B2s is too high in energy to be accessed with our laser sources. Due to the quality of the samples the optical transitions are sufficiently narrow that it is possible to see clearly separated incoming and outgoing resonances. The Raman scattering has been calibrated using the strength of the silicon 520 cm$^{-1}$ Raman peak, taking into account its excitation photon energy dependence [33] and the Fabry-Perot effects associated with the substrates [34] to determine the Raman scattering probability for incoming photons. This calibration allows us to compare the strength of scattering at the different excitonic resonances and between the different materials.

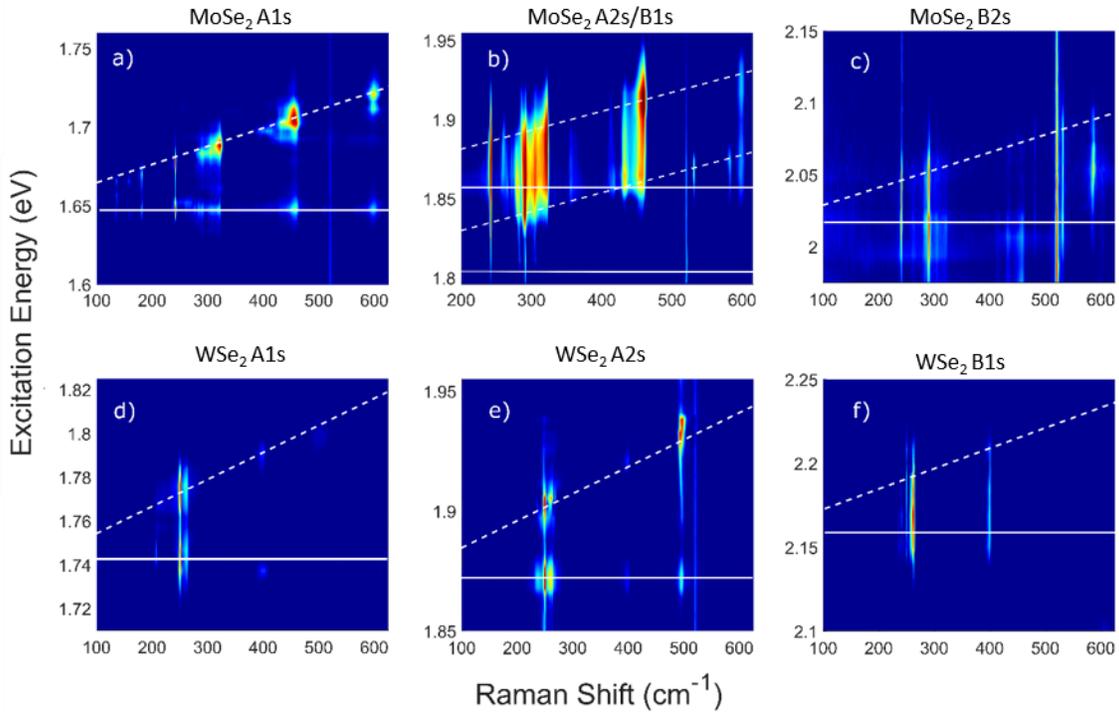

**Figure 1** Color plots of the Resonance Raman spectra for monolayer MoSe$_2$ (a-c) when resonant with the A1s, A2s/B1s and B2s and for WSe$_2$ (d-f) with the A1s, A2s and B1s excitons. The data is plotted using a logarithmic scale for intensity in order to enhance the weaker features. The white lines in each panel show the incident (solid line) and outgoing (dashed lines) resonances expected for each of the observed excitonic states based upon fitting of the $A_1'$ resonances. The exception to this is the WSe$_2$ B1s resonance which instead uses the energy of the B 1s obtained from reflectivity.

In order to discuss the resonance profiles we first need to identify and assign the various Raman peaks. Despite the fact that most of the Raman peaks can be associated with multi-phonon or defect allowed scattering process we obtain a reasonable fit using multiple Lorentzian lineshapes[31]; 18 in MoSe$_2$ and 16 in WSe$_2$. The majority of these peaks have been previously reported and assigned [35–38]. However as shown in the supplementary information (see Tables S3 & S4) there are multiple possible assignments for most of these peaks beyond those already proposed in the literature. Furthermore, comparison of 4 K and room temperature spectra (see Fig S5) demonstrates that a peak 147.3 cm$^{-1}$, previously assigned to LA(M) phonon scattering is associated with a process involving both emission and absorption of phonons. Whilst there is some uncertainty concerning the assignment of several Raman peaks, we can be confident that the 249.3 cm$^{-1}$ peak in WSe$_2$ is due to single phonon scattering with the $A_1'(\Gamma)$ and $E'(\Gamma)$ phonons and that the 241.0 cm$^{-1}$ peak in MoSe$_2$ is also due to single $A_1'(\Gamma)$ phonon scattering.

The resonance profiles of the $A_1'$/$E'$ peaks, presented in Fig 2, should be the simplest to understand as they are due to a single phonon scattering event. We have fitted these resonance profiles using the standard perturbation prediction (see SI) for the Raman scattering probability assuming the underlying process where an incoming photon generates a bright exciton which is scattered by the $A_1'$ phonon to itself or another bright exciton and recombines. Apart from the case of the A2s/B1s MoSe$_2$ transition, for which we use two excitons and allow for interstate scattering, we have assumed that only one exciton is responsible for each resonance. As can be seen in Fig 2, the fits between this simplest theory and the data are remarkably good in all cases apart from the WSe$_2$ B1s resonance. In the case of the WSe$_2$ B1s resonance the profile is clearly asymmetric with the outgoing resonance stronger than the incoming resonance. In addition, there is significant Raman scattering at energies above the resonance but no comparable scattering at the low energy side of the resonance. Both observations can be explained by a series of different models including at least two discrete optically-active states, including the B1s exciton, and, optionally, the onset of a resonance with the C excitonic band whose effects are observed in many optical measurements [37,39] to extend significantly below the energy at which they peak. Whilst it is possible to obtain reasonable fits to the data with plausible models (see SI) it is not possible to select between these models and so we chose not to present a fit.

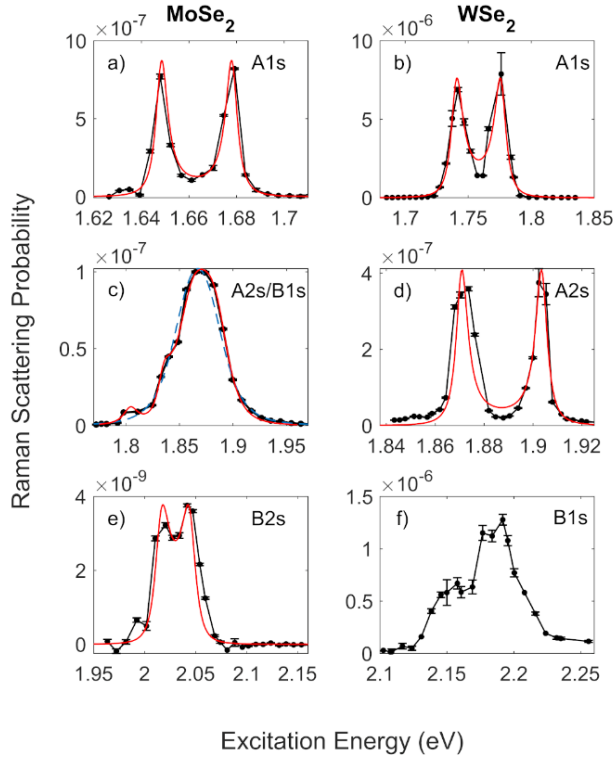

**Figure 2** (a,c,e) Resonance Raman profiles for the MoSe$_2$ A$_1'$ peak are shown when resonant with the A1s, A2s/B1s and B2s excitonic states. (b,d,f) Resonance profiles for the WSe$_2$ A$_1'$ peak are shown when resonant with the A1s, A2s and B1s excitonic states. The red lines show the fitted profiles determined by using a single scattering event model. For all resonance profiles excluding the MoSe$_2$ A2s/B1s and WSe$_2$ B1s the fitted model used assumes a single underlying excitonic states. For the WSe$_2$ A2s/B1s the proximity of these two states requires the use of a two excitonic states model where we have allowed for the possibility of interstate scattering. At the WSe$_2$ B1s resonance we were unable to determine which of a series of possible models was the best fit therefore we do not present a fit. Error bars shown are a standard deviation determined from fitting the Raman spectra.

The excitonic energies, widths and amplitude coefficients obtained from the best fits are presented in Table 1 alongside data obtained from fitting the reflectivity spectra (see SI Figs S5 & S6). The energies determined from the resonance profiles agree to within a few meV with the energies obtained from fits to the PL (A 1s) and reflectivity (A 1s, B 1s) (see SI Tables S2 & S5) and the accepted values in the literature for these transitions[40–43]. In the case of the A2s and B2s transitions in MoSe$_2$ and A 2s in WSe$_2$ there is more limited literature. However, luminescence and reflectivity measurements observe the WSe$_2$ A2s 130 meV above the A1s [12,16,44]. Likewise in MoSe$_2$ optical measurements have observed the A2s and B2s excited states ~ 150 meV above the A1s and B1s exciton. The MoSe$_2$ B2s state has only been reported by Arora et al [45]. Where comparison is possible, the linewidths from the Raman fits are in reasonable agreement with the linewidths obtained from the reflectivity measurements. Apart from the MoSe$_2$ A2s, whose proximity to the much stronger B1s complicates the fitting process, they also follow the pattern, established in the literature, of the 2s states being narrower than their respective 1s states; this observation has been attributed to a reduced radiative coupling. The B resonances are significantly broader than the A resonances, which is attributed to enhanced phonon scattering due to a greater final density of states [24]. As expected from the resonance profiles and spectra shown in Fig 2 & 3, there are significant differences in the amplitude coefficients of the A$_1'$ peak at the different resonances. The leading order perturbation prediction for the scattering strength is proportional to $M_{exciton-phonon}^2 M_{exciton-photon}^4 E_{out}^3 E_{in}$ where the exciton-photon matrix element can be separately constrained by fitting the reflectivity spectra (see SI). A comparison of the values of the ratios of $M^4_{exciton-photon}E^3_{out}E_{in}$ based on the reflectivity fits with the ratio of the Raman amplitude coefficients, WSe$_2$ A2s/A1s 0.008 ± 0.002 and 0.012 ± 0.002, and MoSe$_2$ B1s/A1s 3.0 ± 0.1 and 4.7 ± 0.7, indicates that most, if not all, of the variation of the strength of the Raman scattering can be explained by the dipole matrix element. Therefore, we deduce that exciton-phonon matrix elements are the same for different excitonic states. This is not unexpected, as the underlying deformation potentials are the same [46] and the envelope function should not influence q=0 scattering [6], but hasn't been shown directly in experiments before.

**Table 1** Coefficients from fitting the $A_1'$ resonance Raman profiles and reflectivity spectra for both monolayer MoSe$_2$ and WSe$_2$. From the reflectivity spectra for monolayer MoSe$_2$ (see SI) it is not possible to resolve either the A2s or B2s transitions. The fits to the A2s/B1s resonance in MoSe$_2$ are dominated by the B1s and interstate scattering and as shown by the large error in the A2s amplitude coefficient the resulting parameters should be considered with some caution. The errors shown are a standard deviation determined from the fitting process.

|  | Exciton | Raman | | | Reflectivity | | | |
|---|---|---|---|---|---|---|---|---|
|  |  | $P_a^2$ (eV$^4$) | Γ (meV) | E (meV) | $M^2$ (eV$^2$) | Γ (meV) | E (eV) | $M^2E^4$ |
| WSe$_2$ | A 1s | 306.5±26.4x10$^{-15}$ | 5.6 ± 0.3 | 1740 ± 1 | 1.150 ± 0.024 | 3.3 ± 0.1 | 1736 ± 1 | 12.0 ± 0.5 |
|  | A 2s | 3.8±0.8x10$^{-15}$ | 2.9 ± 0.2 | 1871 ± 1 | 0.103 ± 0.011 | 2.9* | 1866 ± 1 | 0.1 ± 0.03 |
| MoSe$_2$ | A 1s | 7.1±0.5x10$^{-15}$ | 3.0 ± 0.1 | 1648 ± 1 | 0.987 ± 0.015 | 4.4 ± 0.1 | 1649 ± 1 | 7.2 ± 0.2 |
|  | A 2s | 0.3±0.2x10$^{-15}$ | 8.2 ± 1.7 | 1804 ± 1 | - | - | - | - |
|  | B 1s | 33.7±3.0x10$^{-15}$ | 18.9 ± 1.3 | 1858 ± 1 | 1.342 ± 0.32 | 24.0 ± 0.5 | 1855 ± 1 | 21.4 ± 1.1 |
|  | B 2s | 0.3±0.1x10$^{-15}$ | 9.4 ± 0.7 | 2016 ± 1 | - | - | - | - |
|  | A 2s/B 1s | 4.03±0.4x10$^{-15}$ | - | - | - | - | - | - |

Our good quantitative understanding of the $A_1'$ and $E'$ Raman peaks provides a basis on which to look for Raman features unique to the 2s states. In MoSe$_2$ three Raman peaks at 480, 531 and 583cm$^{-1}$, which have not been previously reported, are observed at the A2s/B1s and B2s resonances but not the A1s resonance. At the B2s resonance these are the next strongest peaks after the one phonon peaks. At the A2s/B1s resonance these peaks display an unusual resonance behavior. As can be seen in Fig 1 unlike the other peaks seen at this compound resonance, they are strongly resonant at the B1s incoming resonance energy with no strong resonance at the expected B1s outgoing resonance energy. The shifts of these peaks correspond very closely to the A2s-B1s separation energy ~ 55 meV. Therefore, the natural interpretation of the behavior of these peaks is that they are due to B1s to A2s scattering which is strongly resonant at both the initial and final excitonic states. Whilst these peaks can be assigned to multi-phonon processes associated with both K and M point phonons, they can also be assigned to combinations of Γ point phonons, i.e. 2 $E'(\Gamma)$, $A_1'(\Gamma)+E'(\Gamma)$ and $2A_1'(\Gamma)$ respectively. These Γ point phonons cannot cause intervalley scattering and therefore they cannot resonantly scatter A1s excitons to dark states as there are no lower energy intravalley states. However, in the case of the higher lying excitonic branches resonant scattering to a lower branch exciton with a small wavevector is possible. The failure to observe other combinations of small wavevector optical phonons at the B2s resonance also has a natural explanation. The intravalley scattering process involves sequential single phonon scattering events which should follow selection rules similar to those controlling the strength of single phonon Raman scattering. Thus, we should observe only combinations of those phonons which are strong in the single phonon spectra, i.e. the $A_1'(\Gamma)$ and the $E'(\Gamma)$. Taken together these observations strongly suggest that we have observed for the first time a clear signature of intravalley scattering between different excited Rydberg states, which occurs via the $A_1'(\Gamma)$ and $E'(\Gamma)$ phonons.

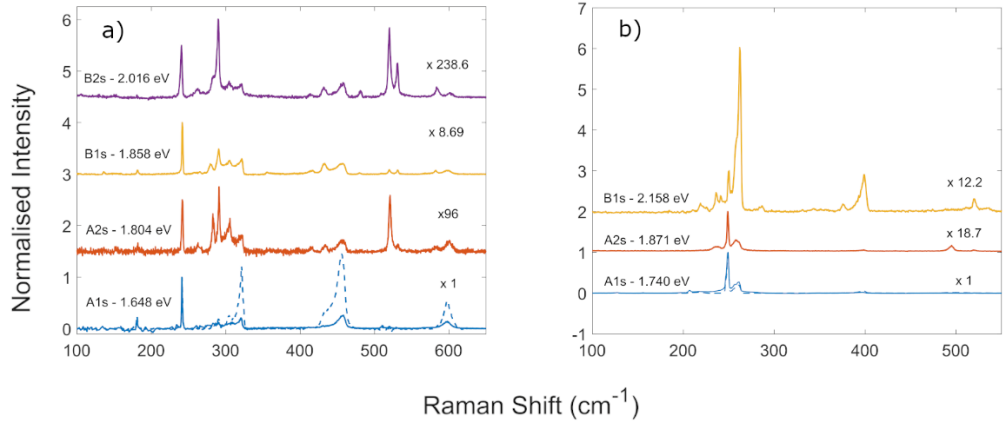

**Figure 3** Raman spectra from MoSe$_2$ and WSe$_2$ monolayers for incoming photon energies at the excitonic energies obtained from fitting the $A_1'$ resonance profiles. At the A1s resonances, a composite Raman spectrum obtained from interpolation of the Raman data is provided showing the intensity of scattering when the outgoing photon is resonant with the A1s resonance (dashed line). The scaling factors are determined from the absolute Raman scattering probability of the $A_1'$ peak scaled relative to the A1s resonance. Note that the A2s and B1s resonances in MoSe$_2$ are close with the B1s scattering significantly stronger and so the A2s spectrum is not solely due to A2s to A2s phonon scattering.

In WSe$_2$ we observe an intriguing dispersive mode, between 490 and 500 cm$^{-1}$, which is only observed at the A2s (Fig 4) resonance. At the incident resonance the Raman peak appears to be a single feature. As the excitation energy is increased from the incoming resonance the Raman peak at first disappears. However, the peak returns as the excitation is increased through a threshold at about 26 meV (210 cm$^{-1}$) greater than the A2s energy. At this point the peak disperses to lower shift with increasing excitation energy in a non-linear manner. At the outgoing resonance the spectrum changes again with three peaks appearing which disperse relatively quickly; two to higher shift and one to lower shift. As can be seen in Fig 4, the higher shift components of the triplet appear to be resonance at an energy higher than the lower with the difference in a manner which cannot be explain as being due to the change in the outgoing resonance energy for the different Raman shifts of the features. The behavior of this feature cannot be explained without one or more dark states with significant features in their dispersion relations, minima or saddle points, at energies comparable with the bright A2s exciton. Unfortunately, no theoretical predictions of the excitonic band structure are available which could be used to identify what these features could be.

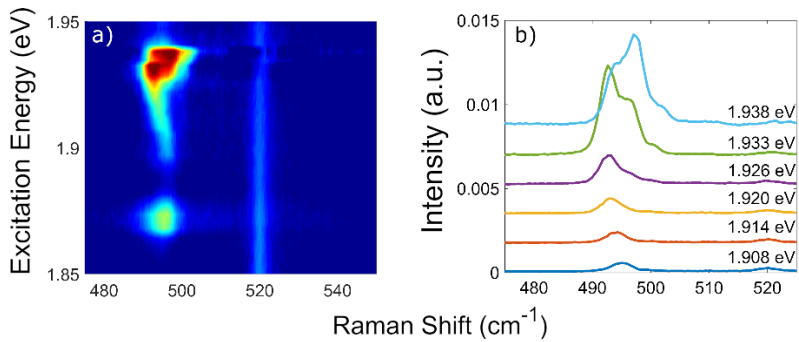

**Figure 4** Color plot (a) and select Raman spectra (b) showing dispersive Raman peaks observed when resonant with the WSe$_2$ A2s exciton. From the color plot (a) we observe a single Raman peak at 495.1 cm$^{-1}$ at the incident resonance. At excitation energies from 1.89 to 1.93 eV this peak demonstrates dispersive behavior with a decrease in the peak frequency. At energies above 1.93 eV we observe three dispersive peaks in the same spectra with frequencies of 492.7, 495.9 and 499.9 cm$^{-1}$ at 1.933 eV. The dispersive behavior of these peaks is complex with the lower

frequency peak continuing to decrease, whereas the two higher frequency peaks are observed to increase in frequency above 1.93 eV.

The data also shows important features not associated with the 2s states. For instance, at the MoSe$_2$ A 1s resonance all the peaks, apart from the A$_1'$, have asymmetric resonance profiles with the outgoing resonance between 4 and 6 times as strong as the incoming resonance (see Fig 3). For comparison Fig 5 shows the resonance profiles for the A$_1'$ peak at both the MoSe$_2$ and WSe$_2$ A1s resonances along with the WSe$_2$ 262 cm$^{-1}$ and MoSe$_2$ 321 cm$^{-1}$ peaks which are both associated with multiphonon processes. It is clear that at the WSe$_2$ A1s resonance the profiles for both the A$_1'$ and 262 cm$^{-1}$ peak are symmetric as expected for a single state resonance, whereas in MoSe$_2$ only the A$_1'$ is symmetric with all other peaks such as a the 321 cm$^{-1}$ peaks demonstrating asymmetric resonance profiles. A natural explanation for the different resonance behavior at the A1s excitons in the two materials is the difference in available excitonic states arising from the conduction band splitting spin splitting. Therefore, we attribute this behavior at the MoSe$_2$ A1s to the sequential two phonon scattering via the intervalley K exciton which is predicted to be 30-40 meV [23,47,48] above the bright A1s exciton. A second clear feature of the data (Fig 3) is that at both the MoSe$_2$ A1s and WSe$_2$ B1s resonances there are two peaks, at 260 and 390 cm$^{-1}$ in WSe$_2$ and 320 and 460 cm$^{-1}$ in MoSe$_2$, which are considerably enhanced; going from approximately one fifth of the A$_1'$ peak to greater than the A$_1'$. It is interesting to note, that the 390 and 460 cm$^{-1}$ peaks are both associated with two other peaks to form triplets. In addition, the 260 and 320 cm$^{-1}$ peaks are both associated with a band of peaks, from 215 to 260 cm$^{-1}$ inWSe$_2$ and 280 to 320 cm$^{-1}$ in MoSe$_2$, and in both cases they are the highest shift peak in the band. Taken together, these observations suggest these peaks are associated with the same underlying phonons shifted in frequency due to the different metal masses. It is not obvious why these peaks aren't resonant at other excitons however one thing the MoSe$_2$ A1s and WSe$_2$ B1s have in common is that they both involve the lower lying conduction band state of the spin split pair, and therefore, their equivalent intervalley excitons involve the higher energy conduction band state. A number of other interesting features of the data including not previously reported dispersive modes observed at the MoSe$_2$ A2s/B 1s and the WSe$_2$ A 1s resonances are discussed in the SI (see Fig S5).

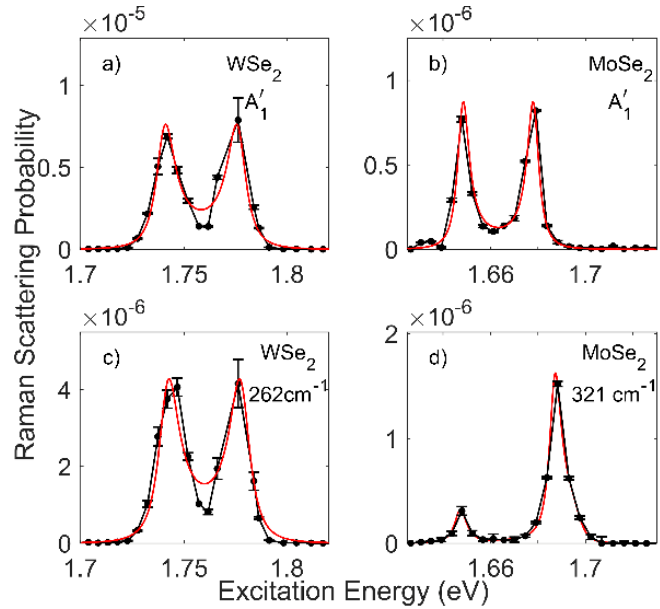

**Figure 5** Resonance Raman profiles for the A$_1'$ peaks in MoSe$_2$ and WSe$_2$ when resonant with the A 1s excitonic state (a-b) along with the profiles of the MoSe$_2$ 321 cm$^{-1}$ and WSe$_2$ 262 cm$^{-1}$ peaks (c-d). For WSe$_2$ both single phonon and multiphonon Raman peaks demonstrate symmetric resonance profiles. In MoSe$_2$ the single phonon A$_1'$ profile is symmetric whilst all other Raman peaks, which have potential assignments to multiphonon/large wavevector scattering processes, demonstrate significant asymmetry between the incident and outgoing resonance peaks (see SI). The contrasting resonance behavior for WSe$_2$ and MoSe$_2$ A1s resonance profiles is attributed to the absence of available large wavevector states at the opposite K valley. Error bars shown are a standard deviation determined from fitting the Raman spectra.

In this paper we present the first observation of Raman scattering from A2s and B2s excitonic states in any TMDCs, along with high resolution resonance Raman results at the A1s and B1s. This allows the differences between phonon scattering processes at different excitonic states to be explored. A quantitative analysis of the absolute Raman scattering probability at various resonances experimentally confirms that the q=0 exciton-phonon scattering matrix elements are the same/ nearly the same for all excitonic states [24]. However, they highlight significant differences between the exciton-phonon scattering of the excitonic states due to the dark states available for scattering. In MoSe$_2$ we observe three peaks at the A2s/B1s and B2s resonances which are clearly associated with intravalley scattering between different Rydberg states. We propose that these peaks can be associated with two phonon scattering by the A$_1'$(Γ) and E'(Γ) phonons and that it is these phonons which dominate intravalley scattering between different Rydberg states. In WSe$_2$ we observe a feature with complex dispersive behavior at the A2s. Whilst it isn't possible to fully explain this peak with the current state of knowledge of the excitonic dispersion relations in these materials, it is clear that the dark excitonic states must have important features at energies close to the A2s resonance.

The data also contains important new features associated with the A1s and B1s resonances. Asymmetric resonance profiles for all of the multi-phonon peaks at MoSe$_2$ A1s state are a direct consequence of the existence of inter-K-valley states with energies a few tens of meV above the Gamma point excitons [49]. In addition, we observe two analogous peaks in each material that are resonant at the A1s in MoSe$_2$ and B1s in WSe$_2$ and which existing literature suggests are not strongly resonant at the C exciton in either material [35,37]. A possible interpretation of the resonance conditions is that it is associated with the different sign spin-orbit coupling in the conduction bands of the two materials.

Finally, our results demonstrate that whilst Raman scattering has already provided significant insight into monolayers and heterostructures of TMDCs there is still a huge amount more than could be extracted from this technique particularly as sample quality improves.

**Methods**

The samples used consist of mechanically exfoliated monolayers of $MoSe_2$ and $WSe_2$ encapsulated between layers of hexagonal boron nitride using a dry transfer technique [50] with the underlying substrate consisting of oxide coated silicon. For low temperature measurements the samples were mounted inside an Oxford Instruments High Resolution liquid helium flow microstat, and unless stated otherwise all experiments were performed at 4 K. All optical measurements were carried out using a back-scattering geometry with a x50 Olympus objective producing a spot size on the sample of 3 µm. Positioning on the sample was achieve a computer controlled 3 axis translation stage and a custom in situ microscope. For all measurements the incident power on the sample was maintained below 100 µW to avoid photo doping and laser heating of the sample [51,52]. The resonance Raman measurements were carried out using a CW Coherent Mira 900 allowing excitation energies from 1.24 to 1.77 eV and a Coherent Cr 599 dye laser using DCM, Rhodamine 6G and Rhodamine 100 laser dyes allowing excitation energies from 1.74 to 2.25 eV. In addition, for PL measurements a 532 nm Coherent Verdi laser was also utilized. The incident polarizations of the lasers were horizontal relative to the optical bench and the Raman scattered light coupled into the spectrometer was analyzed using both horizontal and vertical polarizations. The polarization of the Raman peaks was observed to be strongly co-linear. This allowed unwanted luminescence from the samples to be removed from the Raman spectra by subtraction of the crossed and parallel polarized spectra. Both Pl and Raman spectra were measured using a Princeton Instruments Tri-vista Triple spectrometer equipped with a liquid nitrogen cooled CCD. The Raman peak frequencies were all calibrated using the silicon Raman peak at 520 $cm^{-1}$ as an internal reference. To allow comparison of the Raman scattering on the $MoSe_2$ and $WSe_2$ samples the spectra were calibrated to absolute Raman scattering probability. This required the normalization of the Raman spectra to the 520 $cm^{-1}$ Silicon peak intensity, correction of Fabry-Perot interference effects and calibration using the absolute Raman scattering results of Aggarwal et al [33]. To account for the Fabry-Perot interference we made use of reflectivity spectra measured using a Fianium super continuum source and Ocean optics HR4000 spectrometer. For further detail on the experimental methods, data analysis and Raman calibration please see the supplementary information.

## Supporting Information

The supplementary information contains additional data on both samples presented in the main body of the paper as well as results of measurements on three repeat samples. These include: photoluminescence; reflectivity spectra; details on fitting Raman spectra; Raman peak assignment tables and analysis; reflectivity fitting details; correction of thin film interference effects; Absolute Raman Scattering calibration; Resonance Raman scattering model details, and additional resonance Raman profiles and analysis for all samples.

## Data Availability

The data presented in this paper is openly available from the University of Southampton Repository. DOI:10.5258/SOTON/D1315.

## Author Contributions

The experiments were conceived by D.C.S, L.P.M and X.X. Samples were fabricated by P.R. The experimental measurements were performed by J.V and L.P.M. Data analysis and interpretation was carried out by L.P.M, D.C.S and J.V. The paper was written by D.C.S and L.P.M. All authors discussed the results and commented on the manuscript.

**Corresponding Author**

*D.C.Smith@soton.ac.uk

Present Addresses

**Competing financial interests**

The authors declare no competing financial interests.

**Funding Sources**

Research at the University of Southampton was supported by the Engineering and Physical Science Council of the UK via program grant EP/N035437/1. Both L.P.M and J.V were also supported by EPSRC DTP funding. The work at University of Washington was mainly supported by the Department of Energy, Basic Energy Sciences, Materials Sciences and Engineering Division (DE-SC0018171).